# Generalized BER Performance Analysis for SIC-based Uplink NOMA Systems

Mahmoud AlaaEldin, *Student Member, IEEE*, Emad Alsusa, *Senior Member, IEEE*, Mohammad Al-Jarrah, *Student Member, IEEE*, and Karim G. Seddik, *Senior Member, IEEE*

*Abstract*—Non-orthogonal multiple access (NOMA) is widely recognized for its spectral and energy efficiency, which allows more users to share the network resources more effectively. This paper provides a generalized bit error rate (BER) performance analysis of successive interference cancellation (SIC)-based uplink NOMA systems under Rayleigh fading channels, taking into account error propagation resulting from SIC imperfections. Exact closed-form BER expressions are initially derived for scenarios with 2 and 3 users using quadrature phase shift keying (QPSK) modulation. These expressions are then generalized to encompass any arbitrary rectangular/square M-ary quadrature amplitude modulation (M-QAM) order, number of NOMA users, and number of BS antennas. Additionally, by utilizing the derived closed-form BER expressions, a simple and practically feasible power allocation (PA) technique is devised to minimize the sum bit error rate of the users and optimize the SIC-based NOMA detection at the base-station (BS). The derived closed-form expressions are corroborated through Monte Carlo simulations. It is demonstrated that these expressions can be effective for optimal uplink PA to ensure optimized SIC detection that mitigates error floors. It is also shown that significant performance improvements are achieved regardless of the users' decoding order, making uplink SIC-based NOMA a viable approach.

## I. INTRODUCTION

With the rise of internet-of-things (IoT) and autonomous vehicles, the demand for network resources has been rapidly growing. Hence, the need for spectrally efficient signaling and multiple access (MA) is indispensable. Non-orthogonal multiple access (NOMA) is introduced in the literature as a promising technique for effective spectrum utilization. NOMA can be implemented in the form of power domain NOMA, in which the users are allocated different power levels, and code domain NOMA, in which each user is assigned a distinct code serving as a signature. In the power domain NOMA, which is the focus of this paper, multiple users with different power levels share one resource block and use successive interference cancellation (SIC) to decode the signals. Due to its benefits, NOMA has attracted researchers from academia and industry and is considered an eminent candidate for future generations of cellular networks such as 6G and beyond [1]–[3].

Mahmoud AlaaEldin, Emad Alsusa and Mohammad Al-Jarrah are with the Electrical and Electronic Engineering Department, University of Manchester, Manchester, M13 9PL, UK (e-mail: mahmoud.alaaeldin@manchester.ac.uk; e.alsusa@manchester.ac.uk).

Karim G. Seddik is with the Department of Electronics and Communications Engineering, American University in Cairo, Cairo, Egypt 11835 (e-mail: kseddik@aucegypt.edu).

### A. Related work

The bit error rate (BER)/symbol error rate (SER) analysis of uplink and downlink NOMA were investigated through the past decade under different system parameters like the number of users, modulation orders, and channel type. However, the vast majority of the research body is on downlink NOMA and spans over a wider variety of system parameters. The reason for this is that downlink NOMA systems are more mathematically tractable since all the superimposed NOMA symbols experience the same channel and phase rotation from the base station (BS) to the user equipment (UE), making detection at the UE similar to the detection of quadrature amplitude modulation (QAM). On the other hand, uplink NOMA BER/SER analysis is more challenging since the UEs experience different channels with different phase rotations, making it more difficult to detect the received superimposed NOMA signals at the BS. In the following, we review existing uplink and downlink NOMA error analysis works.

*1) Uplink NOMA error rate analysis:* The primary challenge in analyzing uplink NOMA arises from the need for solving a $K$-fold integration due to the $K$ different channel realizations of the $K$ users affecting the received signal, in order to obtain average performance. To overcome this challenge, different approaches have been introduced. For instance, in [4], the authors utilized the union bound for the outer integration and the exponential bound for the inner integration. They then employed the total probability theorem and Taylor series expansion to calculate the average BER. In another approach, the work in [5] took advantage of the independence of channel realizations and obtained the probability density function (PDF) of the joint channels. Consequently, they derived the exact average BER in a single-integral form and also derived an approximate closed-form expression. However, the approach in [5] only applies to the two-user scenarios and neglects the random phase rotations of the Rayleigh fading channels which is impractical for real scenarios.

In the studies focusing on the uplink NOMA two-user case [6]–[8], closed-form BER expressions have been derived for single input single output (SISO) setups over additive white Gaussian noise (AWGN) channels, considering SIC imperfections. For example, [6] and [7] considered both users employing quadrature phase shift keying (QPSK), while [8] assigned QPSK to the near-user and binary phase shift keying (BPSK) to the far-user to account for channel asymmetry. Furthermore, BER expressions over fading channels have been derived for joint maximum likelihood decoding (JMLD) in [9], [10]. Specifically, [9] considers a SISO Rician fading channel and binary phase shift keying (BPSK), while [10] considers a single input multiple output (SIMO) Rayleigh fading channel and QPSK. Additionally, closed-form union bounds on the SER and BER for a two-user NOMA system with QPSK were derived in [11], valid for both uplink and downlink scenarios. In the uplink, intentional phase rotation can be applied, while in the downlink, the channel can add phase rotation. A multi-carrier under-lay cognitive radio uplink NOMA system was studied in [12]. The authors derived the SER per subcarrier of both the primary and secondary users to evaluate the performance of employing perfect SIC at the receiver under Rayleigh fading channels, and using QPSK modulation. Besides, an upper bound of the BER for an arbitrary number of uplink users with M-phase shift keying (M-PSK) in Rayleigh fading channels was derived in [13] considering JMLD and multiple antennas at the base station. It was found that JMLD overcomes the error floor problem of SIC. The performance of JMLD detection was analyzed in [14] for a $K$-user uplink NOMA system with multiple-antenna BS considering adaptive QAM modulation under Rayleigh fading channels. The authors proposed maximum ratio combining (MRC)-JMLD detection where the BS employs MRC to combine the received signals from all antennas and JMLD for the joint detection of the users' symbols, and the performance of MRC-JMLD detector was evaluated in terms of pairwise error probability (PEP). The results in [14] showed that the utilization of MRC-JMLD effectively eliminates the error floor.

*2) Downlink NOMA error rate analysis:* Downlink NOMA analysis has attracted significant interest within the wireless research community for over a decade covering a wide range of issues from different numbers of users to different modulation orders and fading channels. For example, the authors of [5] studied the BER for a two-user downlink NOMA system under Rayleigh fading channels, assuming QPSK and BPSK modulation schemes are used for the near and far users, respectively. The work in [5] was then extended to consider identical BPSK and QPSK for both users [15]. The authors of [16] derived exact PEP expressions to characterize the performance of two-user and three-user downlink NOMA system using imperfect SIC for ordered Nakagami-$m$ fading channels. They used the obtained PEP expressions to derive an exact union bound on the BER. In [17], the authors derived the exact BER of downlink NOMA systems, considering imperfect SIC decoding over ordered Nakagami-$m$ flat fading channels, for the two-user and three-user scenarios assuming QPSK modulation for all users, whereas they considered the analysis of JMLD in [18]. The authors of [19] derived expressions for the average SER in Rayleigh fading channels considering rectangular $M_n$-QAM, whereas in [20], hexagonal $M_n$-QAM was considered for deriving the average SER under Nakagami–$m$ fading. Exact closed-form BER expressions for imperfect SIC decoding for a two-user downlink NOMA system considering rectangular $M_n$-QAM transmission were derived in [21]. Moreover, the work in [22] quantified the impact of imperfect channel state information (CSI) on the BER performance of a two-user downlink NOMA system assuming BPSK for both users and perfect SIC. The work in [23] presented exact and approximate SER expressions under Rayleigh fading channels for the two-user case considering $\{2, 4\}$-pulse amplitude modulation (PAM) and $\{4, 16\}$-QAM modulations, whereas only $\{2, 4, 8\}$-PAM were considered for the three-user case. Lastly, the authors of [24] derived BER analytical expressions for downlink NOMA with 2 and 3 users in $\alpha - \mu$ fading channels, and generalized the derived expressions for the case of $M$-QAM modulation.

Other works that focused on an arbitrary number of users include [25] which presented PEP analysis for the case of an arbitrary number of users using $M$-QAM modulation. Moreover, in [26], identical $M$-phase shift keying (PSK) modulation orders for all users were assumed. In [27] the authors derived BER expressions with an arbitrary number of users assuming BPSK modulation for all users and perfect SIC under Rayleigh fading channels. They validated the obtained analytical expressions using software defined radio (SDR) experiments. The authors of [28] derived BER expressions for downlink NOMA with an arbitrary number of users, assuming QPSK modulation for all the users. They also presented a gradient descent-based iterative algorithm to find the optimal power allocation of the NOMA users that minimizes the overall BER of the system and validated their results experimentally using SDR. Last but not least, the authors of [29] provided exact and asymptotic BER expressions under Rayleigh fading channels for downlink NOMA systems with arbitrary number of users, arbitrary number of receiving antennas and arbitrary modulation orders, including BPSK and rectangular QAM.

*B. Motivations and Contributions*

Reviewing the uplink NOMA literature has led us to the conclusion that there are several research gaps in the analysis of BER for uplink NOMA systems compared to the downlink NOMA literature. This disparity arises from the fact that the analysis of downlink NOMA is generally more tractable, resulting in closed-form solutions. On the other hand, the uplink NOMA scenario is more challenging.

When the authors of [5] presented the error floor problem in uplink NOMA systems, subsequent research focused primarily on analyzing the performance of JMLD detectors, as they do not suffer from such a problem. However, the complexity of JMLD is extremely high compared to SIC when the number of NOMA users and modulation orders increase. Thus, the practical application of JMLD may be impractical in many real scenarios, particularly those involving high modulation orders and a large number of NOMA users. Therefore, in this manuscript, we investigate the BER performance analysis of low-complexity SIC receivers in uplink NOMA systems under Rayleigh fading channels. We demonstrate that there are ways to mitigate or at least reduce the error floor problem by increasing the number of BS antennas and/or implementing optimized power allocation for the NOMA users.

In this paper, we commence by presenting a comprehensive analytical study on the BER performance of SIC-based uplink NOMA systems. Specifically, we investigate an uplink SIMO NOMA system where single antenna NOMA users transmit data on the same time-frequency resource blocks to a multi-antenna BS over Rayleigh fading channels. By leveraging the multiple antennas at the BS, we combine MRC with SIC decoding, referred to as MRC-SIC, to mitigate the error floor phenomenon and derive exact closed-form BER expressions. The BER expressions are initially derived for the cases of 2 and 3 users using QPSK modulation. Subsequently, the analysis is extended to encompass any rectangular/square $M$-QAM modulation order and any number of NOMA users, while taking into account error propagation resulting from SIC imperfection. Additionally, we propose a simple and efficient power allocation scheme that utilizes the derived expressions to achieve further enhancements in BER. Our results demonstrate a close match between the derived BER expressions and the simulation results. Furthermore, we show that the performance of the proposed MRC-SIC detector becomes comparable to that of JMLD as the number of BS antennas increases. Moreover, our uplink power allocation scheme significantly reduces the BER floors.

The contributions here can be summarized as follows:

- Propose an effective and low complexity uplink NOMA detector, namely MRC-SIC, for decoding NOMA signals in the multi-antenna BS.
- Present a comprehensive BER analytical study of uplink NOMA systems that accounts for SIC imperfections under the assumption of Rayleigh fading channels.
- Account for SIC error propagation to ensure highly accurate closed-form generalized BER expressions for any number of antennas, arbitrary rectangular QAM modulation order, and an arbitrary number of users.
- Propose an effective and low complexity power allocation algorithm for optimzing the SIC receiver's performance and improving the uplink sum BER.
- Validate the presented derivations using extensive simulations and discussions.
- Prove that the BER floor problem in uplink NOMA can be mitigated as the number of antennas at BS increases making the performance of MRC-SIC comparable to JMLD and at significantly less decoding complexity.

*C. Paper organization*

The rest of the paper is organized as follows. In Sec. II, the uplink NOMA-multiple input multiple output (MIMO) system model is presented. Sec. III presents BER analysis for the 2-users and 3-users scenarios when QPSK modulation is adopted by all users. The generalized BER analysis for an arbitrary number of users and square/rectangular QAM modulation orders is presented in Sec. IV. Results and discussion are given in Sec. V, and finally the conclusions are presented in Sec. VI.

## II. SYSTEM MODEL

In this manuscript, we consider an uplink NOMA-SIMO system model that contains a multi-antenna BS with $N$ antennas and $K$ single antenna uplink users, where the $i$th user is denoted by $U_i$. The received signal vector, $\mathbf{y} \in \mathbb{C}^{N \times 1}$, at BS is given as

$$\mathbf{y} = \sum_{i=1}^{K} \sqrt{P_i} \mathbf{h}_i x_i + \mathbf{n}, \quad (1)$$

where $P_i$ denotes the transmit power of $U_i$, and $\mathbf{h}_i$ denotes the wireless channel vector from the $i$th user to the BS. All the elements of $\mathbf{h}_i$ are independent and identically distributed (i.i.d.) complex Gaussian random variables with zero mean and a variance of $\sigma_i^2$ *per one complex dimension*, i.e., $\mathbf{h}_i \sim \mathcal{CN}(0, 2\sigma_i^2 \mathbf{I}_N)$. The additive white Gaussian noise at BS is denoted by $\mathbf{n}$, each element of which is i.i.d. complex Gaussian random variable with zero mean and variance of $\sigma_n^2$ *per complex dimension*, i.e., $\mathbf{n} \sim \mathcal{CN}(0, 2\sigma_n^2 \mathbf{I}_N)$. The term $x_i \in \mathbb{C}$ denotes the transmitted modulation symbol of the $i$th user. The BS is assumed to have perfect channel side information about the channel gain vectors of all the users. Define $\mathbf{g}_i$ as the normalized channel vector of $U_i$, i.e., $\mathbf{g}_i \sim \mathcal{CN}(0, \mathbf{I}_N)$, where $\mathbf{h}_i = \sqrt{2}\sigma_i \mathbf{g}_i$.

Using the MRC-SIC detector, the received signal vector, $\mathbf{y}$ is multiplied first by $\mathbf{h}_1^H$, then the symbol $x_1$ is decoded from $\mathbf{h}_1^H \mathbf{y}$ by treating the other data symbol $x_2$ as noise. The decoded symbol of $U_1$, $\sqrt{P_1}\mathbf{h}_1\widehat{x}_1$, is subtracted from $\mathbf{y}$ to form $\mathbf{y}_2$ then multiplied by $\mathbf{h}_2$ to decode $x_2$, and so on. Therefore, the MRC-SIC detection formula can be expressed as

$$\widehat{x}_k = \arg\min_{x_k \in \mathcal{X}_k} \left| \mathbf{h}_k^H \mathbf{y} - \mathbf{h}_k^H \sum_{i=1}^{k-1} \sqrt{P_i} \mathbf{h}_i \widehat{x}_i - \sqrt{P_k} \|\mathbf{h}_k\|^2 x_k \right|^2, \quad (2)$$

where $\mathcal{X}_k$ is the set of QAM symbol alphabet of the $k$th user, and $\widehat{x}_i$ is the detected symbol of the $i$th user which precedes the $k$th user in the SIC decoding order.

On the other hand, JMLD is considered in this work as a benchmark to compare against. Given perfect CSI knowledge at the BS, the information symbols can be jointly detected using JMLD detection as

$$\{\widehat{x}_1, \widehat{x}_2, \ldots, \widehat{x}_K\} = \arg\min_{x_i \in \mathcal{X}_i} \left\| \mathbf{y} - \sum_{i=1}^{K} \sqrt{P_i} \mathbf{h}_i x_i \right\|^2, \quad (3)$$

where $\{\widehat{x}_1, \widehat{x}_2, \ldots, \widehat{x}_K\}$ are the jointly detected $K$ users' QAM symbols. Although JMLD is the optimal detector for uplink NOMA systems, (3) shows that it suffers from extremely high complexity when the number of users and modulation orders increase. This extremely high complexity makes it impractical to consider in real system implementations.

## III. BER ANALYSIS FOR 2 AND 3 USERS WITH QPSK

In this section, we present detailed BER analysis for all NOMA users in both 2-user and 3-user scenarios, assuming QPSK modulation for all users. The data symbols, $x_1, x_2$, and $x_3$, in (1) are drawn from the set $\mathcal{X}_{\text{QPSK}} = \{s_0 = 1+j, s_1 = -1+j, s_2 = 1-j, s_3 = -1-j\}$. Gray coding is used at all users to map the binary bits to QPSK symbols, where the binary bits are mapped to complex symbols as $00 \rightarrow s_0$, $01 \rightarrow s_1$, $10 \rightarrow s_2$, $11 \rightarrow s_3$. The first and second bits of each QPSK symbol are denoted as $b_{n1}$ and $b_{n2}$, respectively, where $n \in \{1, 2, ..., K\}$ is the index of the $n$th user with $K$ represents the total number of NOMA users.

## A. BER analysis of the 2-user scenario

*1) Analysis of $U_1$:* When decoding $x_1$, MRC is used by multiplying the received signal in (1) by $\mathbf{h}_1^H$ to form $y_1$ as

$$y_1 = \mathbf{h}_1^H \mathbf{y} = \sqrt{P_1}\|\mathbf{h}_1\|^2 x_1 + \sqrt{P_2}\mathbf{h}_1^H \mathbf{h}_2 x_2 + \mathbf{h}_1^H \mathbf{n}, \quad (4)$$

where $\|.\|$ denotes the 2-norm of a vector throughout this paper. By using the fact that $\mathbf{h}_2$ is a complex Gaussian random vector, as well as the noise vector, $\mathbf{n}$, then the interference term of $U_2$ in (4) can be treated as ordinary AWGN given the possible values of $x_2$. Let us define $w_1 \triangleq \sqrt{P_2}\mathbf{h}_1^H \mathbf{h}_2 x_2 + \mathbf{h}_1^H \mathbf{n}$ to denote the combined complex Gaussian noise which affects the decoding process of $x_1$. It should be noted that the distribution of $w_1$ is $\mathcal{CN}(0, 2\sigma_{w_1}^2)$, where $\sigma_{w_1}^2$ represents the variance per complex dimension and it can be calculated as

$$\sigma_{w_1}^2 = P_2\|\mathbf{h}_1\|^2 |x_2|^2 \sigma_2^2 + \|\mathbf{h}_1\|^2 \sigma_n^2. \quad (5)$$

In the QPSK mapping with Gray coding, it can be observed that a bit error event occurs in the first bit when either $s_0$ or $s_1$ moves to the negative side of the imaginary axis affected by imperfect transmission, or when $s_2$ or $s_3$ moves to the positive side of the imaginary axis. Moreover, it can be noticed that these error events occur with equal probability due to the symmetry of Gray coded QPSK. On the other hand, for the second bit, the error event occurs when $s_1$ or $s_3$ moves to the positive side of the real axis, or when $s_0$ or $s_2$ moves to the negative side of the real axis. Therefore, by conditioning on the channel vector of $U_1$, $\mathbf{h}_1$, the average conditional BER of $U_1$ is calculated as the average of the error probability of $b_{11}$ and $b_{12}$ as

$$BER_{U_1|\mathbf{h}_1} = \frac{1}{2}\sum_{i=1}^{2} P_{b_{1i}} = \frac{1}{2}\left[\Pr\left(\mathfrak{Re}(w_1) > \sqrt{P_1}\|\mathbf{h}_1\|^2\right) + \Pr\left(\mathfrak{Im}(w_1) > \sqrt{P_1}\|\mathbf{h}_1\|^2\right)\right] \quad (6)$$

where $P_{b_{1i}}$ is the error probability when detecting $b_{1i}$. Since $w_1$ is circular symmetric Gaussian random variable, it can be concluded that $P_{b_{11}} = P_{b_{12}}$. Hence the average conditional BER of $U_1$ can be expressed as

$$BER_{U_1|\mathbf{g}_1} = Q\left(\sqrt{\frac{P_1\|\mathbf{h}_1\|_2^4}{P_2\|\mathbf{h}_1\|^2|x_2|^2\sigma_2^2 + \|\mathbf{h}_1\|^2\sigma_n^2}}\right)$$
$$= Q\left(\sqrt{\frac{2P_1\sigma_1^2\|\mathbf{g}_1\|^2}{2P_2\sigma_2^2 + \sigma_n^2}}\right), \quad (7)$$

where $Q(.)$ is the $Q$-function, and $|x_2|^2 = 2$, $\forall x_2 \in \mathcal{X}_{\text{QPSK}}$. Let $Z_1 = \|\mathbf{g}_1\|^2$, then the random variable $Z_1$ has an Erlang distribution with rate parameter $\lambda = 1$ whose PDF is

$$f_{Z_1}(z) = \frac{z^{N-1}e^{-z}}{(N-1)!}. \quad (8)$$

BER of $U_1$ can be calculated by averaging $BER_{U_1|Z}$ over the PDF of $Z_1$, and thus $BER_{U_1}$ can be derived as [30]

$$BER_{U_1} = \int_0^\infty Q\left(\sqrt{az}\right) f_{Z_1}(z) dz$$
$$= \frac{1}{2}\left[1 - \sum_{k=0}^{N-1}\binom{2k}{k}\sqrt{\frac{a}{a+2}}\frac{1}{(2a+4)^k}\right], \quad (9)$$

where $a = \frac{2P_1\sigma_1^2}{2P_2\sigma_2^2 + \sigma_n^2}$.

*2) Analysis of $U_2$:* In the SIC detection process, after decoding $x_1$, the detected signal of $U_1$, $\sqrt{P_1}\mathbf{h}_1\widehat{x}_1$, is subtracted from the received vector, $\mathbf{y}$, at the BS to form $\mathbf{y}_2$ as

$$\mathbf{y}_2 = \sqrt{P_2}\mathbf{h}_2 x_2 + \sqrt{P_1}\mathbf{h}_1(x_1 - \widehat{x}_1) + \mathbf{n}. \quad (10)$$

The error propagation term of $U_1$ should be considered as AWGN since $\mathbf{h}_1$ is a complex Gaussian random vector. To decode $x_2$, MRC is used by multiplying the vector, $\mathbf{y}_2$, by $\mathbf{h}_2^H$ to form $y_2$ as

$$y_2 = \mathbf{h}_2^H \mathbf{y}_2 = \sqrt{P_2}\|\mathbf{h}_2\|^2 x_2 + \sqrt{P_1}\mathbf{h}_2^H \mathbf{h}_1(x_1 - \widehat{x}_1) + \mathbf{h}_2^H \mathbf{n}. \quad (11)$$

Thus, following a similar procedure that has been used to derive (7), the conditional BER of $U_2$, given $Z_2 = \|\mathbf{g}_2\|^2$ and the error propagation factor, can be expressed as

$$BER_{U_2|Z_2,d_1} = Q\left(\sqrt{\frac{2P_2\sigma_2^2\|\mathbf{g}_2\|^2}{P_1 d_1^2 \sigma_1^2 + \sigma_n^2}}\right) \quad (12)$$

where $d_1 = |x_1 - \widehat{x}_1|$. Clearly, the BER of $U_2$ depends on $d_1$ which may have different values depending on the accuracy of the decoding process of $x_1$. For instance, assuming $x_1 = s_0$, then we have 3 different possibilities of $d_1$ each of which has a specific probability. The three possible values of $d_1$ can be given as

$$d_1 = \begin{cases} 0, & \widehat{x}_1 = s_0, \\ 2, & \widehat{x}_1 = s_1, s_2, \\ 2\sqrt{2}, & \widehat{x}_1 = s_3. \end{cases} \quad (13)$$

To calculate the average BER of $U_2$, we need to calculate the probability of each possible value that $d$ may take which depends on the pairwise symbol error probability (SEP) of decoding $x_1$. The probability of $d_1 = 0$ can be calculated as

$$\Pr(d_1 = 0|Z_1) = \Pr\left(\mathfrak{Re}(w_1) > \sqrt{P_1}\|\mathbf{h}_1\|^2\right)$$
$$\times \Pr\left(\mathfrak{Im}(w_1) > \sqrt{P_1}\|\mathbf{h}_1\|^2\right) = \left(1 - Q(\sqrt{aZ_1})\right)^2, \quad (14)$$

where $a = \frac{2P_1\sigma_1^2}{2P_2\sigma_2^2 + \sigma_n^2}$. Similarly, the probability of $d_1 = 2$ and $d_1 = 2\sqrt{2}$ can be respectively derived as

$$\Pr(d_1 = 2|Z_1) = 2Q(\sqrt{aZ_1})\left(1 - Q(\sqrt{aZ_1})\right), \quad (15)$$

$$\Pr(d_1 = 2\sqrt{2}|Z_1) = Q(\sqrt{aZ_1})^2. \quad (16)$$

Interestingly, by observing that the QPSK constellation is symmetric, we must have the same values for $d_1$ as in (13) with the same corresponding probabilities for the other possible $x_1$ symbols, i.e., $x_1 = s_2, s_3$, or $s_4$. Now, the 3 pairwise SEPs in (14), (15), and (16) need to be averaged over the PDF of $Z_1$ in (8). Since pairwise SEPs contain $Q^2(.)$ terms, then we can use an exponential approximation for the $Q$-function to derive closed-form expressions for the average unconditional pairwise SEPs. The $Q$-function approximation is given as [31]

$$Q(x) \approx \frac{1}{12}e^{-\frac{x^2}{2}} + \frac{1}{4}e^{-\frac{2}{3}x^2}, \quad x \geq 0. \quad (17)$$

$$P_1^s(a) = 1 + \frac{1}{144(a+1)^N} - \frac{2^N}{6(a+2)^N} - \frac{3^N}{2(2a+3)^N} + \frac{6^N}{24(7a+6)^N} + \frac{3^N}{16(4a+3)^N}, \quad (19)$$

$$P_2^s(a) = \frac{-1}{72(a+1)^N} + \frac{2^N}{6(a+2)^N} - \frac{3^N}{8(4a+3)^N} + \frac{3^N}{2(2a+3)^N} - \frac{6^N}{12(7a+6)^N}, \quad (20)$$

$$P_3^s(a) = \frac{1}{144(a+1)^N} + \frac{6^N}{24(7a+6)^N} + \frac{3^N}{16(4a+3)^N}. \quad (21)$$

Thus, the average unconditional probability of $d = 0$ can be calculated as

$$P_1^s(a) = \int_0^\infty (1 - \frac{1}{12}e^{-\frac{1}{2}az} - \frac{1}{4}e^{-\frac{2}{3}az})\frac{z^{N-1}e^{-z}}{(N-1)!}dz, \quad (18)$$

where this integral can be evaluated using Bernoulli's formula of integration to give the closed-form expression in (19) at the top of page 5. Likewise, the average unconditional probabilities of $d_1 = 0$ and $d_1 = 2\sqrt{2}$ can be calculated using the same way to give closed-form expressions in (20) and (21), respectively, at the top of page 5. Consequently, the average BER of $U_2$, given its channel $\|\mathbf{g}_2\|^2$, can be given as

$$BER_{U_2|Z_2} = \sum_{i=1}^{3} \Pr(d_1 = d_{1i}) BER_{U_2|Z_2,d_1=d_{1i}}$$
$$= \sum_{i=1}^{3} P_i^s(a) Q\left(\sqrt{\frac{2P_2\sigma_2^2 \|\mathbf{g}_2\|^2}{P_1 d_{1i}^2 \sigma_1^2 + \sigma_n^2}}\right), \quad (22)$$

where $d_{11} = 0$, $d_{12} = 2$ and $d_{13} = 2\sqrt{2}$. Finally, by averaging the conditional BER of $U_2$ in (22) over the PDF of $Z_2$, which is identical to the PDF of $Z_1$ in (8), then the average unconditional BER of $U_2$ can be finally found as

$$BER_{U_2} = \frac{1}{2} \sum_{i=1}^{3} P_i^s(a) \left[1 - \sum_{k=0}^{N-1} \binom{2k}{k} \sqrt{\frac{a_i}{a_i+2}} \frac{1}{(2a_i+4)^k}\right], \quad (23)$$

where $a_1 = \frac{2P_2\sigma_2^2}{\sigma_n^2}$, $a_2 = \frac{2P_2\sigma_2^2}{4P_1\sigma_1^2+\sigma_n^2}$, $a_3 = \frac{2P_2\sigma_2^2}{8P_1\sigma_1^2+\sigma_n^2}$.

### B. BER analysis of the 3-users scenario

*1) Analysis of $U_1$ and $U_2$:* Using similar procedures to those used in the 2-user scenario, the BER of both $U_1$ and $U_2$ can be calculated in the same way by considering the additional interference term due to $U_3$'s signal. Therefore, the average BER expressions for $U_1$ and $U_2$ are the same as in the 2-user scenario. The average BER expression for $U_1$ can be given as (9) replacing $a$ with $\gamma$, whereas the average BER expression for $U_2$ can be written as (23) by replacing $a_1$, $a_2$, and $a_3$ with $\gamma_1$, $\gamma_2$, and $\gamma_3$, respectively. The parameters $\gamma$, $\gamma_1$, $\gamma_2$, and $\gamma_3$ are given by

$$\gamma = \frac{2P_1\sigma_1^2}{2P_2\sigma_2^2 + 2P_3\sigma_3^2 + \sigma_n^2}, \quad \gamma_1 = \frac{2P_2\sigma_2^2}{2P_3\sigma_3^2 + \sigma_n^2},$$
$$\gamma_2 = \frac{2P_2\sigma_2^2}{4P_1\sigma_1^2 + 2P_3\sigma_3^2 + \sigma_n^2}, \quad \gamma_3 = \frac{2P_2\sigma_2^2}{8P_1\sigma_1^2 + 2P_3\sigma_3^2 + \sigma_n^2}. \quad (24)$$

*2) Analysis of $U_3$:* After decoding the detected signals of $U_1$ and $U_2$, i.e. $x_1$ and $x_2$, $\sqrt{P_1}\mathbf{h}_1\widehat{x}_1$ and $\sqrt{P_2}\mathbf{h}_2\widehat{x}_2$, are subtracted from the received vector, $\mathbf{y}$, at BS to form $\mathbf{y}_3$ as

$$\mathbf{y}_3 = \sqrt{P_3}\mathbf{h}_3 x_3 + \sqrt{P_1}\mathbf{h}_1(x_1 - \widehat{x}_1) + \sqrt{P_2}\mathbf{h}_2(x_2 - \widehat{x}_2) + \mathbf{n}. \quad (25)$$

The error propagation terms of $U_1$ and $U_2$ are then considered as AWGN as we have done in the 2-user scenario. After that, $\mathbf{y}_3$ is multiplied by $\mathbf{h}_3^H$ for MRC decoding of $x_3$ as

$$y_3 = \mathbf{h}_3^H \mathbf{y}_3 = \sqrt{P_3}\|\mathbf{h}_3\|^2 x_3 + \sqrt{P_1}\mathbf{h}_3^H\mathbf{h}_1(x_1 - \widehat{x}_1) + \sqrt{P_2}\mathbf{h}_3^H\mathbf{h}_2(x_2 - \widehat{x}_2) + \mathbf{h}_3^H\mathbf{n}. \quad (26)$$

Therefore, the conditional BER of $U_3$ can be written as

$$BER_{U_3|Z_3,d_1,d_2} = Q\left(\sqrt{\frac{2P_3\sigma_3^2 \|\mathbf{g}_3\|^2}{P_1 d_1^2 \sigma_1^2 + P_2 d_2^2 \sigma_2^2 + \sigma_n^2}}\right), \quad (27)$$

where $d_1 = |x_1 - \widehat{x}_1|$ and $d_2 = |x_2 - \widehat{x}_2|$. Clearly, the BER of $U_3$ depends on both $d_1$ and $d_2$ which represent the residual errors due to imperfect decoding of $x_1$ and $x_2$, respectively. Similar to the 2-user scenario, both $d_1$ and $d_2$ have 3 different possibilities since QPSK signaling is employed, as given in (13). However, the corresponding probability mass function (PMF)s of $d_1$ and $d_2$ are not independent since the error propagation of $U_1$, $d_1$, does affect the pairwise SEP of $U_2$. Consequently, by using the total probability theorem and considering the dependency between $d_1$ and $d_2$, the average BER of $U_3$ in this case can be calculated as

$$BER_{U_3|Z_3} = \sum_{i=1}^{3} \Pr(d_1 = d_{1i}) \sum_{j=1}^{3} \Pr(d_2 = d_{2j}|d_{1i}) BER_{U_3|Z_3,d_{1i},d_{2j}}$$
$$= \sum_{i=1}^{3} P_i^s(\gamma) \sum_{j=1}^{3} P_j^s(\gamma_i) Q\left(\sqrt{a_{ij}\|\mathbf{g}_3\|^2}\right), \quad (28)$$

where

$$a_{ij} = \frac{2P_3\sigma_3^2}{P_1 d_{1i}^2 \sigma_1^2 + P_2 d_{2j}^2 \sigma_2^2 + \sigma_n^2}, \quad (29)$$

and $d_{11} = d_{21} = 0$, $d_{12} = d_{22} = 2$, $d_{13} = d_{23} = 2\sqrt{2}$, and the functions $P_i^s, \forall i \in \{1, 2, 3\}$ are defined in (19), (20), and (21). Finally, by averaging (28) over the PDF of $Z_3 = \|\mathbf{g}_3\|^2$, which is identical to the PDF of $Z_1$ in (8), the average unconditional BER of $U_3$ can be given as

$$BER_{U_3} = \frac{1}{2} \sum_{i=1}^{3} P_i^s(\gamma) \sum_{j=1}^{3} P_j^s(\gamma_i) \times \left[1 - \sum_{k=0}^{N-1} \binom{2k}{k} \sqrt{\frac{a_{ij}}{a_{ij}+2}} \frac{1}{(2a_{ij}+4)^k}\right]. \quad (30)$$

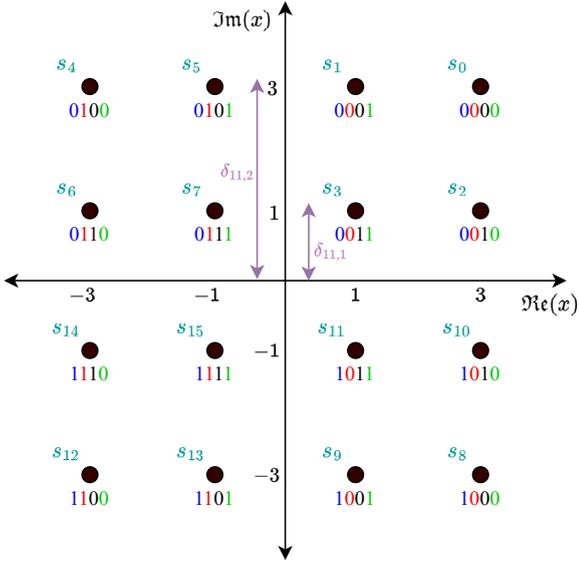

Figure 1: Bit mapping of 16-QAM constellation using Gray coding

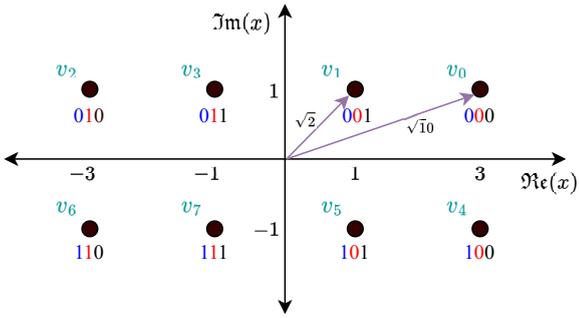

Figure 2: Bit mapping of 8-QAM constellation using Gray coding

## IV. BER ANALYSIS FOR THE K-USER SCENARIO WITH HIGHER ORDER QAM MODULATION SCHEMES

In this section, we consider the analysis of a general $K$-user scenario where every user uses any arbitrary rectangular $M$-QAM modulation. For better illustration and presentation, we first adopt an example where the system contains 3 users assuming 16-QAM modulation for $U_1$, and 8-QAM for $U_2$ and $U_3$. After that, we generalize the BER expressions to consider arbitrary number of users and any QAM modulation order.

For the adopted example, Gray coding is used for all the users to map the binary data to QAM symbols, where each 16-QAM symbol is the mapping of a codeword with 4-bit size, while each 8-QAM symbol is the mapping of a 3-bit codeword, as shown in Fig. 1 and Fig. 2. The $i$th bit of each QAM symbol is denoted as $b_{ni}$ where $n$ is the user index.

### A. Analysis of $U_1$

$U_1$ is the first user in the SIC order, which uses 16-QAM modulation. The received signal vector, $\mathbf{y}$, in (1) is multiplied by $\mathbf{h}_1^H$ for MRC-based decoding of $x_1$ as

$$y_1 = \sqrt{P_1}\|\mathbf{h}_1\|^2 x_1 + \sqrt{P_2}\mathbf{h}_1^H\mathbf{h}_2 x_2 + \sqrt{P_3}\mathbf{h}_1^H\mathbf{h}_3 x_3 + \mathbf{h}_1^H\mathbf{n}. \quad (31)$$

By considering the interference terms of $U_2$ and $U_3$ in (31) as AWGN given the possible values of $x_2$ and $x_3$, the variance of the real or imaginary part of the total effective noise, $w_1 \triangleq \sqrt{P_2}\mathbf{h}_1^H\mathbf{h}_2 x_2 + \sqrt{P_3}\mathbf{h}_1^H\mathbf{h}_3 x_3 + \mathbf{h}_1^H\mathbf{n}$, affecting $U_1$ is given as

$$\sigma_{w_1}^2 = P_2\|\mathbf{h}_1\|^2 |x_2|^2 \sigma_2^2 + P_3\|\mathbf{h}_1\|^2 |x_3|^2 \sigma_3^2 + \|\mathbf{h}_1\|^2 \sigma_n^2. \quad (32)$$

To evaluate the average BER of $U_1$, we derive the probability of error for $b_{11}$, $b_{12}$, $b_{13}$, and $b_{14}$ then we take the average of the 4 bit error probabilities as

$$BER_{U_1|\mathbf{h}_1} = \frac{1}{4}\sum_{i=1}^{4} P_{b_{1i}}. \quad (33)$$

For $b_{11}$, the constellation can be divided into two regions based on the decision boundary, which is the $x$-axis, as shown in Fig. 1. In this case, the decision boundary and the constellation point can be at a distance of $\delta_{11,1} = 1$ or $\delta_{11,2} = 3$, and the error event for $b_{11}$ depends on the transmitted 16-QAM symbol. For example, if $s_0, s_1, s_5,$ or $s_4$ is transmitted, then $\Pr(\widehat{b}_{11} \neq b_{11}) = \Pr\left(\mathfrak{Im}(w_1) > 3\sqrt{P_1}\|\mathbf{h}_1\|^2\right)$, and similarly, if $s_2, s_3, s_7,$ or $s_6$ is transmitted, then $\Pr(\widehat{b}_{11} \neq b_{11}) = \Pr\left(\mathfrak{Im}(w_1) > \sqrt{P_1}\|\mathbf{h}_1\|^2\right)$. The case for $b_{12}$ is similar to $b_{11}$, except that the decision boundary is the $y$-axis, consequently, $P_{b_{11}} = P_{b_{12}}$ and they are given as

$$P_{b_{11}} = P_{b_{12}} = \frac{1}{2}\left[Q\left(\sqrt{\frac{P_1\|\mathbf{h}_1\|^2}{\sigma_{tot1}^2}}\right) + Q\left(\sqrt{\frac{9P_1\|\mathbf{h}_1\|^2}{\sigma_{tot1}^2}}\right)\right], \quad (34)$$

where $\sigma_{tot1}^2$ is given as

$$\sigma_{tot1}^2 = \sigma_n^2 + |x_3|^2 P_3 \sigma_3^2 + |x_2|^2 P_2 \sigma_2^2. \quad (35)$$

On the other hand, Fig. 1 shows that, if one of the symbols in the first or fourth rows is transmitted, then $b_{13}$ will be detected incorrectly if the transmitted symbol is detected as one of the symbols in the second or third rows of the constellation diagram, i.e., $\{s_2, s_3, s_7, s_6, s_{10}, s_{11}, s_{15}, s_{14}\}$. This corresponds to the case where $\mathfrak{Im}(y_1)$ is in the interval $\left[-2\sqrt{P_1}\|\mathbf{h}_1\|^2, 2\sqrt{P_1}\|\mathbf{h}_1\|^2\right]$, with a probability of occurrence equals to $\Pr(\widehat{b}_{13} \neq b_{13}) = \Pr\left(5\sqrt{P_1}\|\mathbf{h}_1\|^2 > \mathfrak{Im}(w_1) > \sqrt{P_1}\|\mathbf{h}_1\|^2\right)$. However, if one of the symbols in the second or third rows is transmitted, then $b_{13}$ will be detected incorrectly if the transmitted symbol is detected as one of the symbols in the first or fourth rows of the constellation diagram, i.e., $\{s_0, s_1, s_5, s_4, s_8, s_9, s_{13}, s_{12}\}$. This corresponds to the case where $\mathfrak{Im}(y_1)$ is in the intervals $\left[-\infty, -2\sqrt{P_1}\|\mathbf{h}_1\|^2\right]$ or $\left[2\sqrt{P_1}\|\mathbf{h}_1\|^2, \infty\right]$, with a probability of occurrence equals to $\Pr(\widehat{b}_{13} \neq b_{13}) = \Pr\left((\mathfrak{Im}(w_1) > \sqrt{P_1}\|\mathbf{h}_1\|^2) \cup (\mathfrak{Im}(w_1) < -3\sqrt{P_1}\|\mathbf{h}_1\|^2)\right)$.

Similarly, same calculations can be made to derive the probability of detection error for $b_{14}$, replacing $\mathfrak{Im}(y_1)$ with $\Re(y_1)$ and the horizontal decision boundaries with vertical ones. Consequently, $P_{b_{13}} = P_{b_{14}}$ and they can be given as

$$P_{b_{13}} = P_{b_{14}} = \frac{1}{2}\left[2Q\left(\sqrt{\frac{P_1\|\mathbf{h}_1\|^2}{\sigma_{tot1}^2}}\right) + Q\left(\sqrt{\frac{9P_1\|\mathbf{h}_1\|^2}{\sigma_{tot1}^2}}\right) - Q\left(\sqrt{\frac{25P_1\|\mathbf{h}_1\|^2}{\sigma_{tot1}^2}}\right)\right], \quad (36)$$

Therefore, by applying (33), the overall average conditional BER of $U_1$ in this case can be given as

$$BER_{U_1|\mathbf{h}_1,x_1,x_2} = \frac{1}{4}\left[3Q\left(\sqrt{\frac{P_1\|\mathbf{h}_1\|^2}{\sigma_{tot1}^2}}\right) + 2Q\left(\sqrt{\frac{9P_1\|\mathbf{h}_1\|^2}{\sigma_{tot1}^2}}\right) - Q\left(\sqrt{\frac{25P_1\|\mathbf{h}_1\|^2}{\sigma_{tot1}^2}}\right)\right], \quad (37)$$

Since $\sigma_{tot1}^2$ is a function of $|x_2|^2$ and $|x_3|^2$ as in (35), and both can be equal to 2 or 10 in the case of 8-QAM, as shown in Fig. 2, the overall unconditional BER of $U_1$ is given as

$$BER_{U_1} = \frac{1}{4}\sum_{|x_2|^2 \in \mathcal{S}}\sum_{|x_3|^2 \in \mathcal{S}}\frac{1}{4}\left[3f(c_{11}) + 2f(c_{12}) - f(c_{13})\right], \quad (38)$$

where $\mathcal{S} = \{2, 10\}$, $c_{11} = \frac{2P_1\sigma_1^2}{\sigma_{tot1}^2}$, $c_{12} = \frac{18P_1\sigma_1^2}{\sigma_{tot1}^2}$, $c_{13} = \frac{50P_1\sigma_1^2}{\sigma_{tot1}^2}$, and the function $f(x)$ is given as

$$f(x) = \frac{1}{2}\left[1 - \sum_{k=0}^{N-1}\binom{2k}{k}\sqrt{\frac{x}{x+2}}\frac{1}{(2x+4)^k}\right]. \quad (39)$$

### B. Analysis of $U_2$

$U_2$ is the second user in the SIC order, which uses 8-QAM. To begin, the detected signal of $U_1$, $\sqrt{P_1}\mathbf{h}_1\widehat{x}_1$ is subtracted from the received signal, $\mathbf{y}$, then the resultant vector is multiplied by $\mathbf{h}_2^H$ for MRC-based decoding of $x_2$ as

$$y_2 = \sqrt{P_2}\|\mathbf{h}_2\|^2 x_2 + \sqrt{P_1}\mathbf{h}_2^H\mathbf{h}_1(x_1 - \widehat{x}_1) + \sqrt{P_3}\mathbf{h}_2^H\mathbf{h}_3 x_3 + \mathbf{h}_2^H\mathbf{n}. \quad (40)$$

Considering the interference term of $U_3$ and the error propagation term of $U_1$ in (40) as AWGN given the possible values of $|x_3|$ and $|x_1 - \widehat{x}_1|$, the variance of the real or the imaginary part of the total effective noise affecting $U_2$ can be given as

$$\sigma_{w_2}^2 = P_1\|\mathbf{h}_2\|^2|x_1 - \widehat{x}_1|^2\sigma_1^2 + P_3\|\mathbf{h}_2\|^2|x_3|^2\sigma_3^2 + \|\mathbf{h}_2\|^2\sigma_n^2. \quad (41)$$

The error propagation term, $|x_1 - \widehat{x}_1|$, depends on the average pairwise SEP of $U_1$ which must be calculated to derive the average BER of $U_2$. However, the pairwise SEP of $U_1$ itself depends on the values of $|x_2|$ and $|x_3|$ which control the variance of the total effective noise affecting $U_1$. Therefore, we have different pairwise SEP of $U_1$ depending on the possible combinations of $|x_2|$ and $|x_3|$ values, which makes the calculation of BER of $U_2$ follows the tree in Fig. 3, where $P_i^{s1}$ is the pairwise SEP of $U_1$, as illustrated next.

Since both $x_2$ and $x_3$ are drawn from the 8-QAM constellation shown in Fig. 2, then we have 4 possible combinations of the pair $(|x_2|, |x_3|)$ as each of them can only have the values $\sqrt{2}$ or $\sqrt{10}$. Given that $x_1 = s_0$, then $\widehat{x}_1$ could be any of the 16-QAM points in Fig. 1, including $s_0$, which makes 16 possibilities for $|x_1 - \widehat{x}_1|$. The same happens if $x_1$ took the values $s_1$, $s_2$ or $s_3$, making a total of 64 different possibilities for $|x_1 - \widehat{x}_1|$, which illustrates the tree diagram in Fig. 3. We do not count the possibilities when $x_1$ takes a symbol in one of the other 3 quads since it gives repetitive values for $|x_1 - \widehat{x}_1|$ due to the symmetry of the constellation among the 4 quads.

We have 64 different pairwise SEP values which corresponds to the 64 possible error distances of $|x_1 - \widehat{x}_1|$. The average pairwise SEP of $U_1$ given that $x_1 = s_k$ can be given as in (43), where $f_{Z_1}$ and $\sigma_{tot1}$ are given in (8) and (35), respectively. The factor, $\Pr(x_1 = s_k) = \frac{1}{4}$, in (43) is the prior probability of $x_1$ taking one of the 4 symbols in the first quad or their similar points in the other 3 quads. Since (43) involves products of the Q-function, then the exponential approximation of the Q-function can be used to solve the integral in (43), where the approximation can be given as

$$Q(x) \approx \begin{cases} \frac{1}{12}e^{-\frac{x^2}{2}} + \frac{1}{4}e^{-\frac{2}{3}x^2}, & x \geq 0, \\ 1 - \frac{1}{12}e^{-\frac{x^2}{2}} - \frac{1}{4}e^{-\frac{2}{3}x^2}, & x < 0. \end{cases} \quad (42)$$

For each possible combination of $(|x2|, |x3|)$, we have different pairwise SEP vector that corresponds to the 64 values of $|x_1 - \widehat{x}_1|$. Thus, by denoting that the possible 64 error distances of $|x_1 - \widehat{x}_1|$ as $d_1, d_2, \ldots, d_{64}$, and following the tree in Fig. 3, the average BER of $U_2$ can be given as

$$BER_{U_2|\mathbf{h}_2} = \frac{1}{4}\sum_{\mathcal{A}_j \in \mathcal{A}}\sum_{d_i} P_i^{s1}(\mathcal{A}_j)\Pr(\text{error}|\mathcal{A}_j, d_i, \mathbf{h}_2), \quad (44)$$

where $P_i^{s1}(\mathcal{A}_j)$ is given in (43), and $\mathcal{A}$ is the set of all possible events $\mathcal{A} = \{\mathcal{A}_1, \mathcal{A}_2, \mathcal{A}_3, \mathcal{A}_4\}$, as defined in Fig. 3.

Now, we calculate $\Pr(\text{error}|\mathcal{A}_j, d_i, \mathbf{h}_2)$ in (44) which represents the conditional BER of $U_2$ given $\mathcal{A}_j$, $d_i$ and $\mathbf{h}_2$. To evaluate the average BER of $U_2$, we derive the probability of error for $b_{21}$, $b_{22}$, and $b_{23}$, then we take the average of the 3 bit error probabilities as

$$\Pr(\text{error}|\mathcal{A}_j, d_i, \mathbf{h}_2) = \frac{1}{3}\sum_{i=1}^{3} P_{b_{2i}}. \quad (45)$$

We have two cases for the probability calculation in (45), one when $\mathcal{A}_j = \mathcal{A}_1$ or $\mathcal{A}_2$, and another when $\mathcal{A}_j = \mathcal{A}_3$ or $\mathcal{A}_4$. In the case of $\mathcal{A}_j = \mathcal{A}_1$ or $\mathcal{A}_2$, the value $|x_2| = \sqrt{2}$, and hence, $x_2$ is restricted to one of the 4 constellation symbols $v_1, v_3, v_5, v_7$ as shown in Fig. 2. Consequently, we calculate the error probability of the three bits of $U_2$, $P_{b_{21}|\mathcal{A}_{1,2}}$, $P_{b_{22}|\mathcal{A}_{1,2}}$, and $P_{b_{23}|\mathcal{A}_{1,2}}$, given only $x_2 \in \{v_1, v_3, v_5, v_7\}$. For $b_{21}$, the constellation can be divided into two regions based on the decision boundary, which is the $x$-axis, as shown in Fig. 2. In this case, the decision boundary and the constellation points can be at a distance of 1. Then $\Pr(\widehat{b}_{21} \neq b_{21}) = \Pr\left(\Im(w_2) > \sqrt{P_2}\|\mathbf{h}_2\|^2\right)$. Consequently, $P_{b_{21}|\mathcal{A}_{1,2}}$ can be given as

$$P_{b_{21}|\mathcal{A}_{1,2}} = Q\left(\sqrt{\frac{P_2\|\mathbf{h}_2\|^2}{\sigma_{tot2}^2}}\right), \quad (46)$$

where $\sigma_{tot2}^2$ is given by

$$\sigma_{tot2}^2 = \sigma_n^2 + |x_3|^2 P_3\sigma_3^2 + |d_i|^2 P_1\sigma_1^2. \quad (47)$$

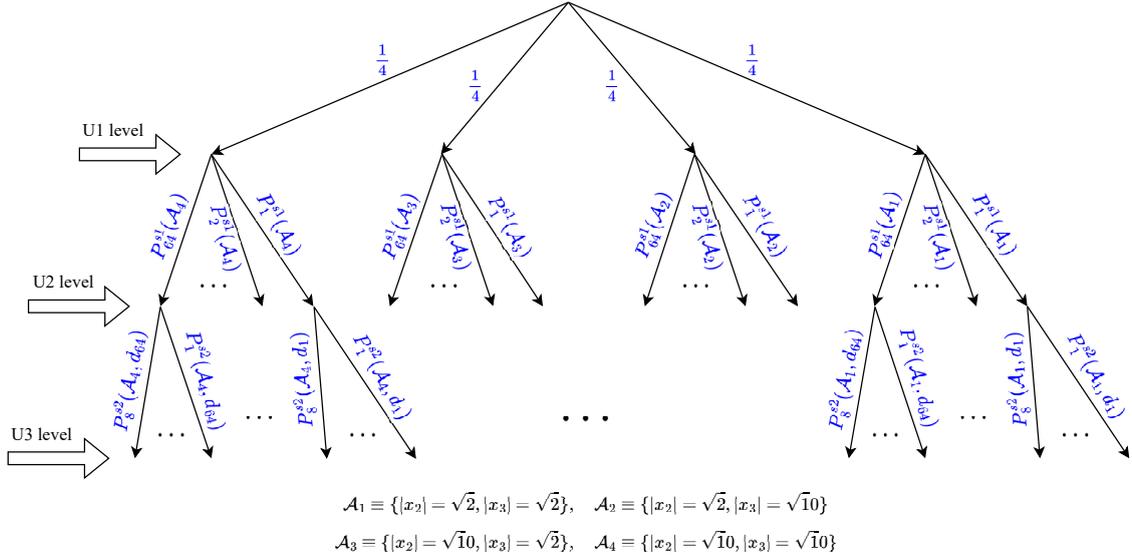

$\mathcal{A}_1 \equiv \{|x_2| = \sqrt{2}, |x_3| = \sqrt{2}\}, \quad \mathcal{A}_2 \equiv \{|x_2| = \sqrt{2}, |x_3| = \sqrt{10}\}$
$\mathcal{A}_3 \equiv \{|x_2| = \sqrt{10}, |x_3| = \sqrt{2}\}, \quad \mathcal{A}_4 \equiv \{|x_2| = \sqrt{10}, |x_3| = \sqrt{10}\}$

Figure 3: 16-8-8 QAM BER tree

For $b_{22}$, the constellation can be divided into two regions based on the decision boundary, as shown in Fig. 2. In this case, the decision boundary and the constellation points can be at a distance of 1. Then $\Pr(\widehat{b}_{22} \neq b_{22}) = \Pr\left(\mathfrak{Re}(w_2) > \sqrt{P_2}\|\mathbf{h}_2\|^2\right)$. Consequently, $P_{b_{22}|\mathcal{A}_{1,2}}$ can be given as

$$P_{b_{22}|\mathcal{A}_{1,2}} = Q\left(\sqrt{\frac{P_2\|\mathbf{h}_2\|^2}{\sigma_{tot2}^2}}\right). \tag{48}$$

However, for $b_{23}$, it will be detected incorrectly if the transmitted symbol is detected as one of the symbols in the first or fourth columns of the constellation diagram, i.e., $\{v_2, v_6, v_0, v_4\}$. This corresponds to the case where $\mathfrak{Re}(y_2)$ is in the intervals $\left[-\infty, -2\sqrt{P_2}\|\mathbf{h}_2\|^2\right]$ or $\left[2\sqrt{P_2}\|\mathbf{h}_2\|^2, \infty\right]$ with a probability of occurrence equals to $\Pr(\widehat{b}_{23} \neq b_{23}|\mathcal{A}_{1,2}) = \Pr\left(\left(\mathfrak{Re}(w_2) > \sqrt{P_2}\|\mathbf{h}_2\|^2\right) \cup \left(\mathfrak{Re}(w_2) < -3\sqrt{P_2}\|\mathbf{h}_2\|^2\right)\right)$. Hence, $P_{b_{23}|\mathcal{A}_{1,2}}$ can be given as

$$P_{b_{23}|\mathcal{A}_{1,2}} = Q\left(\sqrt{\frac{P_2\|\mathbf{h}_2\|^2}{\sigma_{tot2}^2}}\right) + Q\left(\sqrt{\frac{9P_2\|\mathbf{h}_2\|^2}{\sigma_{tot2}^2}}\right). \tag{49}$$

Therefore, by substituting the derived values of $P_{b_{21}|\mathcal{A}_{1,2}}$, $P_{b_{22}|\mathcal{A}_{1,2}}$, and $P_{b_{23}|\mathcal{A}_{1,2}}$ in (45), then the overall conditional BER of $U_2$, in the case of $\mathcal{A}_j = \mathcal{A}_1$ or $\mathcal{A}_2$, can be written as

$$\Pr(\text{error}|\mathcal{A}_{1,2}, d_i, \mathbf{h}_2) = \frac{1}{3}\left[3Q\left(\sqrt{\frac{P_2\|\mathbf{h}_2\|^2}{\sigma_{tot2}^2}}\right) + Q\left(\sqrt{\frac{9P_2\|\mathbf{h}_2\|^2}{\sigma_{tot2}^2}}\right)\right]. \tag{50}$$

On the other hand, in the case that $\mathcal{A}_j = \mathcal{A}_3, \mathcal{A}_4$, the value $|x_2| = \sqrt{10}$, and hence, $x_2$ is restricted to one of the 4 constellation points $v_0, v_2, v_4, v_6$ as shown in Fig. 2. In this case, the probability of detecting $b_{21}$ in error is the same as the previous case, i.e., $P_{b_{21}|\mathcal{A}_{3,4}} = P_{b_{21}|\mathcal{A}_{1,2}}$, which is given in (46). However, for $b_{22}$, the constellation can be divided into two regions based on the decision boundary, which is the $y$-axis, as in the previous case. Different from the previous case, the decision boundary and the constellation points are at a distance of 3. Then $\Pr(\widehat{b}_{22} \neq b_{22}) = \Pr\left(\mathfrak{Re}(w_2) > 3\sqrt{P_2}\|\mathbf{h}_2\|^2\right)$. Consequently, $P_{b_{22}|\mathcal{A}_{3,4}}$ can be given as

$$P_{b_{22}|\mathcal{A}_{3,4}} = Q\left(\sqrt{\frac{9P_2\|\mathbf{h}_2\|^2}{\sigma_{tot2}^2}}\right). \tag{51}$$

For $b_{23}$, it will be detected incorrectly if the transmitted symbol is detected as one of the symbols in the second or third columns of the constellation diagram, i.e., $\{v_1, v_3, v_5, v_7\}$. This corresponds to the case $\mathfrak{Re}(y_2)$ is in the interval $\left[-2\sqrt{P_2}\|\mathbf{h}_2\|^2, 2\sqrt{P_2}\|\mathbf{h}_2\|^2\right]$ with a prob-

$$P_i^{sk}(\mathcal{A}, d_1) = \int_0^\infty \Pr(x_k = s_w)\left[Q\left(\sqrt{\frac{2P_k\sigma_k^2(\alpha_m - \mathfrak{Re}(s_w))^2 z}{\sigma_{totk}^2}}\right) - Q\left(\sqrt{\frac{2P_k\sigma_k^2(\alpha_{m+1} - \mathfrak{Re}(s_w))^2 z}{\sigma_{totk}^2}}\right)\right] \times$$
$$\left[Q\left(\sqrt{\frac{2P_k\sigma_k^2(\beta_n - \mathfrak{Im}(s_w))^2 z}{\sigma_{totk}^2}}\right) - Q\left(\sqrt{\frac{2P_k\sigma_k^2(\beta_{n+1} - \mathfrak{Im}(s_w))^2 z}{\sigma_{totk}^2}}\right)\right] f_{Z_k}(z)dz,$$
$$\text{If } k = 1 \Longrightarrow \forall m, n \in \{1, 2, 3, 4\}, \quad \forall w \in \{0, 1, 2, 3\}, \quad \alpha_m, \beta_n \in \{-\infty, -2, 0, 2, \infty\},$$
$$\text{If } k = 2 \Longrightarrow \forall m \in \{1, 2, 3, 4\}, \forall n \in \{1, 2\}, \quad \forall w \in \{0, 1\}, \quad \alpha_m \in \{-\infty, -2, 0, 2, \infty\}, \beta_n \in \{-\infty, 0, \infty\}. \tag{43}$$

ability of occurrence equals to $\Pr(\widehat{b}_{23} \neq b_{23}|\mathcal{A}_{3,4}) = \Pr\left(5\sqrt{P_2}\|\mathbf{h}_2\|^2 > \Re(w_2) > \sqrt{P_2}\|\mathbf{h}_2\|^2\right)$. Hence,

$$P_{b_{23}|\mathcal{A}_{3,4}} = Q\left(\sqrt{\frac{P_2\|\mathbf{h}_2\|^2}{\sigma_{tot2}^2}}\right) - Q\left(\sqrt{\frac{25P_2\|\mathbf{h}_2\|^2}{\sigma_{tot2}^2}}\right). \quad (52)$$

By substituting the derived values of $P_{b_{21}|\mathcal{A}_{3,4}}$, $P_{b_{22}|\mathcal{A}_{3,4}}$, and $P_{b_{23}|\mathcal{A}_{3,4}}$ in (45), then the overall conditional BER of $U_2$, in the case of $\mathcal{A}_j = \mathcal{A}_3$ or $\mathcal{A}_4$, can be written as

$$\Pr(\text{error}|\mathcal{A}_{3,4}, d_i, \mathbf{h}_2) = \frac{1}{3}\left[2Q\left(\sqrt{\frac{P_2\|\mathbf{h}_2\|^2}{\sigma_{tot2}^2}}\right) + Q\left(\sqrt{\frac{9P_2\|\mathbf{h}_2\|^2}{\sigma_{tot2}^2}}\right) - Q\left(\sqrt{\frac{25P_2\|\mathbf{h}_2\|^2}{\sigma_{tot2}^2}}\right)\right]. \quad (53)$$

Finally, the average unconditional BER of $U_2$, averaged over the channel norm, $\|\mathbf{h}_2\|^2$, can be given as

$$BER_{U_2} = \frac{1}{4}\sum_{(|x_2|,|x_3|)\in\mathcal{A}}\sum_{d_i} P_i^{s1}(\mathcal{A}_j)f_{\mathcal{A}_j}, \quad (54)$$

where $f_{\mathcal{A}_j}$ is the average of the expressions in (50) and (53) over the Erlang distribution of the channel norm, $\|\mathbf{h}_2\|^2$, and it can be expressed as

$$f_{\mathcal{A}_j} = \begin{cases} \frac{1}{3}[3f(c_{21}) + f(c_{22})], & j=1,2, \\ \frac{1}{3}[2f(c_{21}) + f(c_{22}) - f(c_{23})], & j=3,4, \end{cases} \quad (55)$$

where function $f$ is defined in (39), $c_{21} = \frac{2P_2\sigma_2^2}{\sigma_{tot2}^2}$, $c_{22} = \frac{18P_2\sigma_2^2}{\sigma_{tot2}^2}$, $c_{23} = \frac{50P_2\sigma_2^2}{\sigma_{tot2}^2}$.

*C. Analysis of $U_3$*

After decoding both $x_1$ and $x_2$, the detected signals of $U_1$ and $U_2$, $\sqrt{P_1}\mathbf{h}_1\widehat{x}_1$ and $\sqrt{P_2}\mathbf{h}_2\widehat{x}_2$, respectively, are subtracted from the received vector, $\mathbf{y}$, at the BS to form $\mathbf{y}_3$ as

$$\mathbf{y}_3 = \sqrt{P_3}\mathbf{h}_3 x_3 + \sqrt{P_1}\mathbf{h}_1(x_1 - \widehat{x}_1) + \sqrt{P_2}\mathbf{h}_2(x_2 - \widehat{x}_2) + \mathbf{n}. \quad (56)$$

Similar to the QPSK case, the error propagation terms of $U_1$ and $U_2$ are considered as AWGN, and $\mathbf{y}_3$ is multiplied by $\mathbf{h}_3^H$ for MRC decoding of $x_3$ as

$$y_3 = \mathbf{h}_3^H \mathbf{y}_3 = \sqrt{P_3}\|\mathbf{h}_3\|^2 x_3 + \sqrt{P_1}\mathbf{h}_3^H\mathbf{h}_1(x_1 - \widehat{x}_1) + \sqrt{P_2}\mathbf{h}_3^H\mathbf{h}_2(x_2 - \widehat{x}_2) + \mathbf{h}_3^H\mathbf{n}. \quad (57)$$

Hence, the total effective noise affecting $U_3$ can be given as

$$\sigma_{tot3}^2 = \sigma_n^2 + |d_{2j}|^2 P_2\sigma_2^2 + |d_{1i}|^2 P_1\sigma_1^2, \quad (58)$$

where $d_{1i}$ and $d_{2j}$ are the possible values of the quantities, $|x_1 - \widehat{x}_1|$ and $|x_2 - \widehat{x}_2|$, respectively. Since $U_3$ uses 8-QAM as $U_2$, then its conditional bit error probability is similar to $U_2$ expressions in (50) and (53). Therefore, the conditional bit error probability of $U_3$ can be expressed as

$$\Pr(\text{error}|\mathcal{A}_{1,2}, d_{1i}, d_{2j}, \mathbf{h}_3) = \frac{1}{3}\left[3Q\left(\sqrt{\frac{P_3\|\mathbf{h}_3\|^2}{\sigma_{tot3}^2}}\right) + Q\left(\sqrt{\frac{9P_3\|\mathbf{h}_3\|^2}{\sigma_{tot3}^2}}\right)\right], \quad (59)$$

and,

$$\Pr(\text{error}|\mathcal{A}_{3,4}, d_{1i}, d_{2j}, \mathbf{h}_3) = \frac{1}{3}\left[2Q\left(\sqrt{\frac{P_3\|\mathbf{h}_3\|^2}{\sigma_{tot3}^2}}\right) + Q\left(\sqrt{\frac{9P_3\|\mathbf{h}_3\|^2}{\sigma_{tot3}^2}}\right) - Q\left(\sqrt{\frac{25P_3\|\mathbf{h}_3\|^2}{\sigma_{tot3}^2}}\right)\right]. \quad (60)$$

Clearly, the BER of $U_3$ depends on $d_{1i}$ and $d_{2j}$ which represent the symbol errors of $U_1$ and $U_2$ during the decoding process of $x_1$ and $x_2$, respectively. However, the corresponding PMFs of $d_{1i}$ and $d_{2j}$ are not independent since the error propagation of $U_1$, $d_{1i}$, does affect the pairwise SEP of $U_2$. Hence, BER of $U_3$ in this case shall follow the tree diagram in Fig. 3 to consider the dependency between $d_{1i}$ and $d_{2j}$ as

$$BER_{U_3|\mathbf{h}_3} = \frac{1}{4}\sum_{\mathcal{A}_l\in\mathcal{A}}\sum_{d_{1i}} P_i^{s1}(\mathcal{A}_l)\sum_{d_{2j}} P_j^{s2}(\mathcal{A}_l)\times \Pr(\text{error}|\mathcal{A}_l, d_{1i}, d_{2j}, \mathbf{h}_3), \quad (61)$$

where $P_i^{s1}$ and $P_j^{s2}$ are given in (43). $d_{1i}$ has 64 possible values while $d_{2j}$ have only 8 values given the value of $x_2$ and $x_3$. Finally, by averaging (61) over the channel norm, $\|\mathbf{h}_3\|^2$, the average unconditional BER of $U_3$ can be given as

$$BER_{U_3} = \frac{1}{4}\sum_{(|x_2|,|x_3|)\in\mathcal{A}}\sum_{d_{1i}} P_i^{s1}(\mathcal{A}_l)\sum_{d_{2j}} P_j^{s2}(\mathcal{A}_l) f_{\mathcal{A}_l}, \quad (62)$$

where $f_{\mathcal{A}_l}$ is the average of the expressions in (59) and (60) over the Erlang distribution of the channel norm, $\|\mathbf{h}_3\|^2$, and it can be given as

$$f_{\mathcal{A}_l} = \begin{cases} \frac{1}{3}[3f(c_{31}) + f(c_{32})], & l=1,2, \\ \frac{1}{3}[2f(c_{31}) + f(c_{32}) - f(c_{33})], & l=3,4, \end{cases} \quad (63)$$

where the function $f$ is defined in (39), $c_{31} = \frac{2P_3\sigma_3^2}{\sigma_{tot3}^2}$, $c_{32} = \frac{18P_3\sigma_3^2}{\sigma_{tot3}^2}$, $c_{33} = \frac{50P_3\sigma_3^2}{\sigma_{tot3}^2}$.

*D. Generalization of the BER expressions*

In this subsection, we generalize the presented BER expressions to include an arbitrary number of users and any square/rectangular QAM modulation order. Firstly, we start by deriving the BER expressions for an arbitrary number of users assuming QPSK modulation is used by all users. Afterwards, the obtained BER expressions are generalized to include any higher QAM modulation order.

In the QPSK modulation case, the BER of $U_1$ can be written as in (9) by substituting the parameter $a$ as

$$a = \frac{2P_1\sigma_1^2}{2\sum_{i=2}^{K} P_i\sigma_i^2 + \sigma_n^2}, \quad (64)$$

where $K$ is the total number of users. By using the total probability theorem and by generalizing (28), we can deduce that the BER of the $k$th user, $\forall k \geq 2$, in the SIC order can be given as

$$BER_{U_k} = \sum_{i_1=1}^{3} P_{i_1}^s(a) \sum_{i_2=1}^{3} P_{i_2}^s(a_{i_1}) \cdots \sum_{i_{k-1}=1}^{3} P_{i_{k-1}}^s(a_{i_1,\dots,i_{k-2}}) \times \left[1 - \sum_{n=0}^{N-1}\binom{2n}{n}\sqrt{\frac{a_{i_1,\dots,i_{k-1}}}{a_{i_1,\dots,i_{k-1}}+2}}\frac{1}{(2a_{i_1,\dots,i_{k-1}}+4)^n}\right]. \quad (65)$$

where $P_i^s$ is given in (19), (20) and (21), and

$$a_{i_1,\ldots,i_{k-1}} = \frac{2P_k\sigma_k^2}{\sum_{j=1}^{k-1} P_j d_{j,i_j}^2 \sigma_j^2 + \sum_{j=k+1}^{K} 2P_j\sigma_j^2 + \sigma_n^2}. \quad (66)$$

On the other hand, when the NOMA users use any other QAM modulation order, the pairwise SEPs and the bit error probabilities differ from the QPSK case. The total average BER of the $(k+1)$th user in this case can be represented as

$$\begin{aligned}
BER_{U_{k+1}} &= \frac{1}{|\mathcal{A}|} \sum_{\mathcal{A}_l \in \mathcal{A}} \sum_{d_{1,i_1}} \Pr(\widehat{x}_1 = s_{i_1}|\mathcal{A}_l) \cdots \\
&\quad \sum_{d_{k,i_k}} \Pr(\widehat{x}_k = s_{i_k}|\mathcal{A}_l, \mathbf{d}_{k-1}) P_{e|\mathcal{A}_l,\mathbf{d}_k} \\
&= \frac{1}{|\mathcal{A}|} \sum_{\mathcal{A}_l \in \mathcal{A}} \sum_{d_{1,i_1}} P_{i_1}^{s1}(\mathcal{A}_l) \cdots \sum_{d_{k,i_k}} P_{i_k}^{sk}(\mathcal{A}_l, \mathbf{d}_{k-1}) P_{e|\mathcal{A}_l,\mathbf{d}_k},
\end{aligned}$$
(67)

where $|\mathcal{A}|$ is the cardinality of the set $\mathcal{A}$, i.e., the number of elements in the set, and $\mathbf{d}_k = [d_{1,i_1},\ldots,d_{k,i_k}]^T$ is a random vector that contains the symbol error distances of the users up to $U_k$. It should be noted that $\mathcal{A}$ is the set of all possible combinations of the first quadrant of QAM symbols that the NOMA users can take from their alphabets. Only the first quadrant of QAM symbols are taken into account due to the symmetry with the other quadrants. The terms $P_{i_k}^{sk}$ represent the pairwise SEPs which are given in (72) for any rectangular QAM order. The term, $s_{k,l}$, in (72) is the modulation symbol of user $k$ in $\mathcal{A}_l$, whereas $\sigma_{k_{tot}}^2$ is the effective noise variance affecting user $k$ and it can be given as

$$\sigma_{k_{tot}}^2 = \sum_{j=1}^{k} P_j d_{j,i_j}^2 \sigma_j^2 + \sum_{j=k+2}^{K} P_j |x_j|^2 \sigma_j^2 + \sigma_n^2. \quad (68)$$

$P_{e|\mathcal{A}_l,\mathbf{d}_k}$ in (67) is the average conditional BER of $U_{k+1}$ given its transmitted symbol $s_{k+1,l}$ in $\mathcal{A}_l$. The bit error probability for a specific QAM symbol can be calculated in the same way as in the illustrated 16-8-8 QAM example. However, for a general QAM order assuming Gray coding, the conditional bit error probability can be approximated by only taking into account the neighbouring regions of the given QAM symbol. In this case, the conditional BER given any QAM symbol except for the corner points and the edge points of the constellation can be well approximated as

$$\begin{aligned}
P_{e|\mathcal{A}_l,\mathbf{d}_k} &\simeq \frac{1}{\log_2(M)} \int_0^\infty 4Q\left(\sqrt{\frac{2P_{k+1}\sigma_{k+1}^2 z}{\sigma_{k+1_{tot}}^2}}\right) f_{Z_{k+1}}(z) dz \\
&= \frac{4f(\eta)}{\log_2(M)},
\end{aligned}$$
(69)

where $\eta = \frac{2P_{k+1}\sigma_{k+1}^2}{\sigma_{k+1_{tot}}^2}$, and $M$ is the QAM modulation order. Similarly, the BER expression given one of the 4 corner points of the constellation can be given as

$$P_{e|\mathcal{A}_l,\mathbf{d}_k} \simeq \frac{2f(\eta)}{\log_2(M)}, \quad (70)$$

whereas the BER given one of the edge points of the constellation can be given as

$$P_{e|\mathcal{A}_l,\mathbf{d}_k} \simeq \frac{3f(\eta)}{\log_2(M)}. \quad (71)$$

*E. Average users' BER minimization based power allocation*

In this subsection, we discuss the uplink power optimization of the users so that the average BER of all users at BS is minimized. The optimization search for this problem is clearer and more robust to be done in the log-log domain where both the power values to be optimized, $P_k$, and the cost function are in dB. This simple transformation leads to smoother numerical optimization and faster convergence of the gradient descent-based optimization tools. Thus, the uplink power allocation problem can be formulated as

$$\min_{P_1,P_2,\ldots,P_K} \quad 10\log_{10}\left(\sum_{k=1}^{K} BER_{U_k}\left(10^{\frac{P_1}{10}},\ldots,10^{\frac{P_K}{10}}\right)\right) \quad (73a)$$

$$\text{s.t.} \quad P_k \leq P_{\text{dBm}}^{\max}, \quad \forall k, \quad (73b)$$

where $P_{\text{dBm}}^{\max}$ represents the maximum available uplink transmit power in dBm. The cost function in (73a) is the sum of the average BER expressions of the users derived above. E.g., in the QPSK case, $BER_{U_k}$ can take the general form for an arbitrary number of users in (65), whereas for the general QAM modulation case, $BER_{U_k}$ can be expressed using (67).

## V. SIMULATION RESULTS

In this section, we consider a study where the number of uplink NOMA users is fixed at 3. We plot the BER performance of the three users against the transmit power under various values of $N$ and channel gain ratios. The ratio between the channel envelopes of the users are varied by changing the channels standard deviation, $\sigma_1$, $\sigma_2$, and $\sigma_3$, where the ratio is defined as $\rho \triangleq \frac{\sigma_1}{\sigma_2} = \frac{\sigma_2}{\sigma_3}$. The Gaussian noise variance in the real or imaginary dimensions is set to unity. The simulated BER is plotted up to $\text{BER} = 10^{-7}$ due to limitation on the number of Monte Carlo runs, whereas the analytical solution is plotted to much smaller values.

Fig. 4 shows the achievable BER for a 3-user system using 4-QAM with a single antenna BS. The figure also shows the impact of using optimized users' power allocation, discussed in Sec. IV-E, on the achievable BER. The channels are subject to Rayleigh fading with standard deviations of $\sigma_1 = 10$, $\sigma_2 = 10/4$, $\sigma_3 = 10/16$, for users $U_1$, $U_2$, and $U_3$, respectively. The figure clearly shows a perfect match between the simulated BER and the derivations. Also, it can be noticed from the figure that without power allocation, the BER of all users suffers from a significant error floor. For instance, $U_1$, $U_2$, and $U_3$ suffer from error floors of $2.8 \times 10^{-3}$, $4 \times 10^{-3}$ and $4 \times 10^{-3}$, respectively. However, the optimal power allocation algorithm proposed in this paper shows some robustness against the error floor behaviour, as it manages to remove all error floors. Even though the rate at which the BER decreases as a function of the transmit power, the results provide an indication that the introduced power allocation algorithm is beneficial for uplink NOMA when a single antenna BS is deployed and SIC detection is applied. It can be also observed from the figure that SIC and MLD detection provide comparable performance, however, as SNR increases they start diverging where the BER of SIC converges to an error floor while that of MLD keeps decreasing. Also, although the power allocation algorithm can remove the error floor of SIC, there is still a significant performance gap in favour of MLD. e.g., at BER = $10^{-4}$, MLD provides almost 18 dB SNR gain compared to SIC with power allocation.

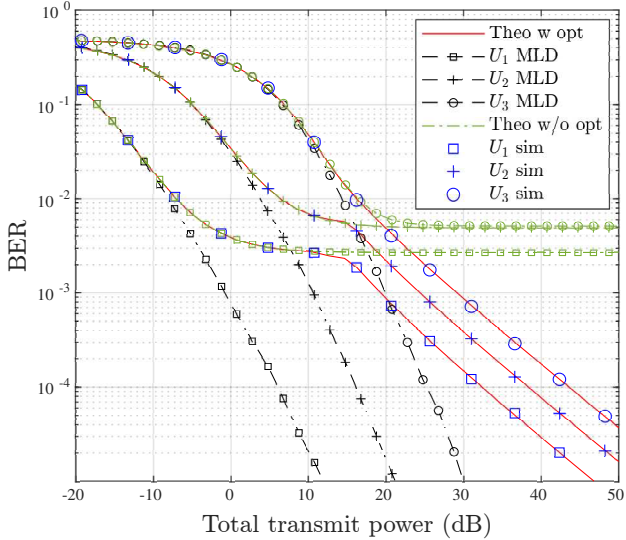

Figure 4: Performance of the optimized uplink NOMA system using 4-QAM modulation, $N = 2$, $\sigma_1 = 10$, $\sigma_2 = 10/4$, $\sigma_3 = 10/16$

Fig. 5 compares the achievable BER for different channel envelope gain ratios, namely, $\rho = \{4, 10\}$. Three users employing 4-QAM are used in this figure, and the BS is equipped with $N = 4$ antennas. The figure confirms the correctness of the derived formulas for the BER of SIC as the simulation and analytical results are identical. As can be also seen from the figure, the BER of the first user is better than the second user which comes in the second place, whereas the third user comes in the final place with the worst BER. It can be also observed from the figure that when the ratio $\rho$ increases, the error floor will decrease resulting in a better BER at high SNR, even though low values of $\rho$ are more preferred at low SNR for users $U_2$ and $U_3$. This behaviour is referred to the fact that increasing $\rho$ implies that the channels of $U_2$ and $U_3$ become worse, and thus the BER at low transmit powers is worse than that of small $\rho$. On the other hand, at large values of the transmit power and $\rho$, the performance is dominated by the accuracy of SIC which leads to BER enhancements until the BER saturates again at higher transmit power values. As can be also observed from the figure, the gap between SIC and MLD is less than that in Fig. 4 which is due to increasing the number of receiving antennas $N$. Interestingly, when $\rho = 10$, the performance gap between MLD and SIC is negligible. Therefore, SIC might be more preferred under these operating conditions as it provides comparable BER to MLD with far less computational complexity.

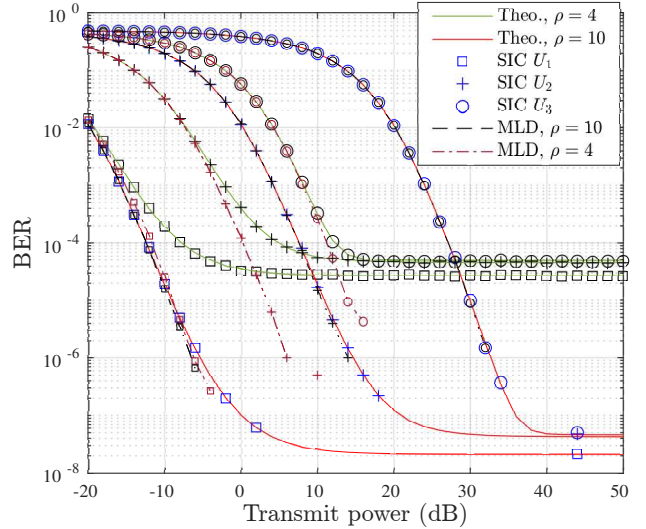

Figure 5: Performance of uplink NOMA system using 4-QAM modulation, $N = 4$, $(\sigma_1, \sigma_2, \sigma_3) = (10, 1, 0.1)$, $(10, 10/4, 10/16)$

Fig. 6 shows the BER under the same system parameters used in Fig. 5, except for the number of BS antennas which is set to $N = 10$ here. By comparing the BER curves in this figure with their counterparts in figure with Fig. 5, it can be observed that there is a significant positive impact of increasing $N$ on the achievable BER. More specifically, the BER is significantly improved by increasing $N$ from 4 to 10. For example, in the case of $N = 4$, an SNR of 0 dB is required for $U_1$ when $\rho = 10$ to achieve a BER of $10^{-7}$, whereas the same BER is achieved at SNR of $-15$ dB when $N = 10$. In addition, the figure clearly shows that increasing the number of antennas to $N = 10$ manages to remove the error floor for all users, or reduce it to a level that is much less than $10^{-8}$. A perfect match between simulation and theoretical BER for SIC in this figure is also observed. As can be seen, the performance gap between MLD and SIC is negligible for all cases, which makes SIC preferable as it is less complex than MLD.

Fig. 7 illustrates the BER of SIC based uplink NOMA using high modulation orders assuming $\rho = 4$, i.e., $(\sigma_1, \sigma_2, \sigma_3) = (10, 10/4, 10/16)$, under different number of antennas. The modulation order considered is 16-QAM for $U_1$ and 8-QAM for both $U_2$ and $U_3$. It is worth highlighting that the near user, i.e., $U_1$, typically has a strong channel and thus it could transmit with higher rate or high modulation orders. On the other hand, the far users usually suffer from relatively bad

$$P_{i_k}^{sk}(\mathcal{A}_l, \mathbf{d}_{k-1}) = \int_0^\infty \left[ Q\left(\sqrt{\frac{2P_k \sigma_k^2 (\alpha_m - \Re(s_{k,l}))^2 z}{\sigma_{k_{tot}}^2}}\right) - Q\left(\sqrt{\frac{2P_k \sigma_k^2 (\alpha_{m+1} - \Re(s_{k,l}))^2 z}{\sigma_{k_{tot}}^2}}\right) \right] \times$$
$$\left[ Q\left(\sqrt{\frac{2P_1 \sigma_1^2 (\beta_n - \Im(s_{k,l}))^2 z}{\sigma_{k_{tot}}^2}}\right) - Q\left(\sqrt{\frac{2P_1 \sigma_1^2 (\beta_{n+1} - \Im(s_{k,l}))^2 z}{\sigma_{k_{tot}}^2}}\right) \right] f_{Z_k}(z) dz,$$
$$\forall m, n \in \{1, 2, \cdots, \sqrt{M}\}, \quad \alpha_m, \beta_n \in \{-\infty, -\sqrt{M}+2, -\sqrt{M}+4, \cdots, \sqrt{M}-2, \infty\}. \tag{72}$$

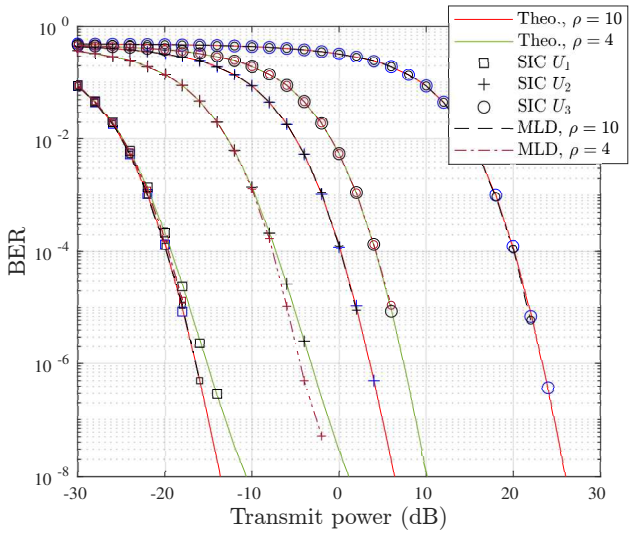

Figure 6: Performance of uplink NOMA system using 4-QAM modulation, $N = 10$, $(\sigma_1, \sigma_2, \sigma_3) = (10, 1, 0.1), (10, 10/4, 10/16)$

BER and thus lower modulation orders are preferable. In both figures, there is a perfect match between the simulation and analysis for the BER for the SIC detector, which again confirms that the derived equations in this paper are accurate for various system parameters. The figure clearly shows the positive impact of $N$ on BER, for example, about 17 dB gain can be obtained for $U_1$ by increasing $N$ from 14 to 20 at BER $= 10^{-7}$. Interestingly, as can be seen from these subplots, the BER of $U_1$ is the lowest even though it employs a higher modulation order that is 16-QAM. This is due to the fact that the channel conditions for $U_1$ is excellent compared to the other two users. Moreover, as can be seen from Fig. 7, the BER suffers from error floor even though $N = 8$ antennas is employed, for example, the floor when $\rho = 4$ is in the range of $10^{-4}$ for $U_2$ and $U_3$. By comparing these results with those in Fig. 5, it can be seen that a higher error floor is incurred in Fig. 7 which is due to the larger constellation size.

Fig. 8 illustrates the BER of SIC-based uplink system employing different modulation orders at the users. As can be observed from the Fig. 8.a, with a modulation order of 64, $U_1$ manages to provide better performance than $U_2$ and $U_3$ even though they employ lower modulation orders, i.e., 8 and $\{2, 4\}$ for users $U_2$ and $U_3$, respectively. This is due to the fact that the channel condition of $U_1$ is far better than the other two users. As can be also seen from Fig. 8.a, for which $N = 8$, by increasing the modulation for $U_1$ from 64 to 256, its BER is degraded by about 6 dB at BER $= 10^{-4}$. However, the BER of $U_1$ is still comparable to the BER of the other two users, that is, it is lower than $U_3$ but slightly higher than $U_2$ for a wide range of SNR. The results in these two subplots clearly show that $U_1$ can have a higher communication rate, i.e., higher modulation order, with comparable BER to far users. Interestingly, the error floor suffered by all users in Fig. 8.a, is removed and the BER is significantly enhanced when $N$ is increased to 20, as can be seen from Fig. 8.b.

## VI. Conclusions

This work considered uplink SIMO-NOMA systems where multiple single-antenna users transmit data streams to a multi-antenna BS. A BER analysis is conducted for a generalized system model that considers an arbitrary number of users, modulation order, and number of receiving antennas. Moreover, a simple power allocation algorithm based on the derived expressions is proposed. The performance of the optimized SIC receiver is compared to that of the optimal JMLD.

The results demonstrate that the proposed power allocation algorithm effectively eliminates the error floors experienced in uplink NOMA. A comparison between the SIC detector and JMLD reveals that the latter yields significant performance improvement when the number of antennas is relatively small. However, as the number of antennas increases, the performance gap between the two approaches diminishes. It is important to note that while the JMLD detector provides better performance, it suffers from significantly high complexity, which further escalates with higher modulation orders and number of users. Consequently, JMLD may not be practical for real-world implementation when employing high modulation orders. The study also shows that increasing the number of receiving antennas at the BS substantially reduces the SIC BER. Additionally, the results demonstrate that by achieving comparable BER for all users, near users can transmit data with higher modulation orders, resulting in higher data rates.

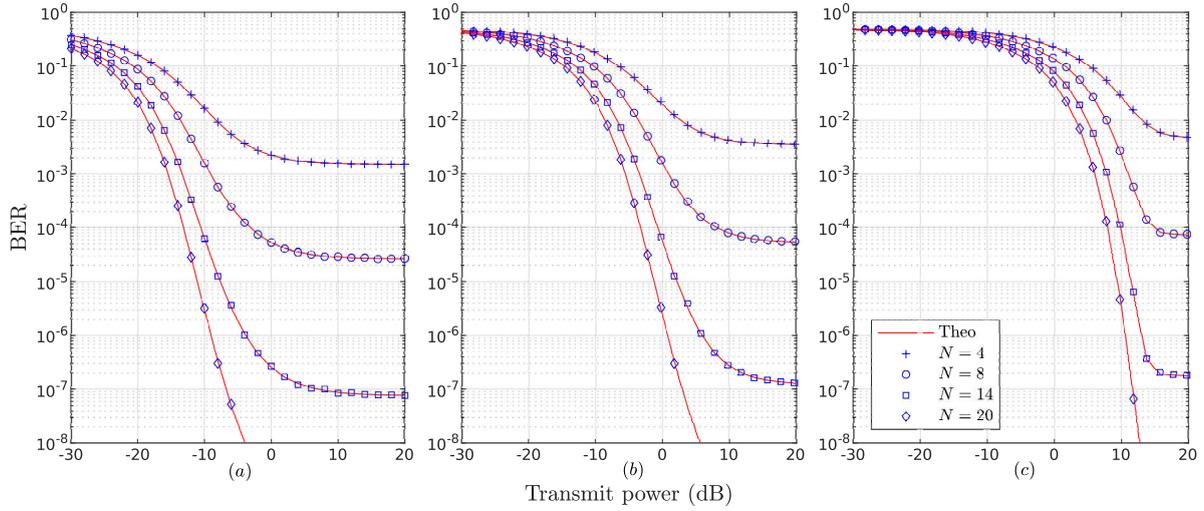

Figure 7: Performance of SIC based uplink NOMA with 16-QAM for $U_1$, and 8-QAM for $U_2$ and $U_3$, $(\sigma_1, \sigma_2, \sigma_3) = (10, 10/4, 10/16)$

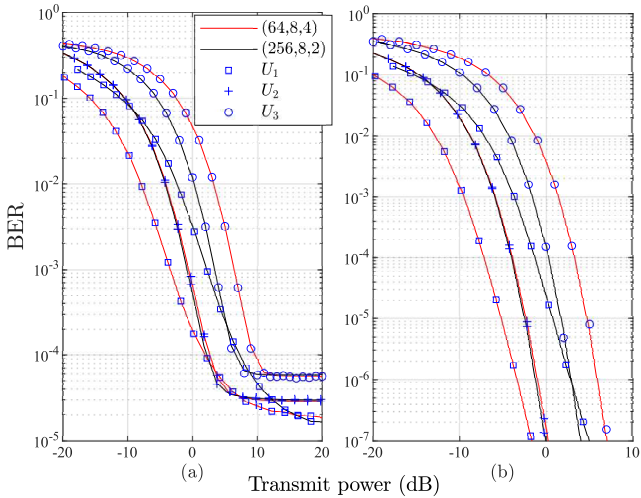

Figure 8: SIC BER Comparisons under different data rates for the cases of (a) $N = 8$, (b) $N = 20$, $(\sigma_1, \sigma_2, \sigma_3) = (10, 10/4, 10/16)$.